\documentclass[]{spie}  

\usepackage[numbers]{natbib}
\usepackage{graphicx} 
\usepackage{url}
\usepackage{amsmath, amssymb}

\usepackage{xcolor}
\usepackage[colorlinks=true, allcolors=blue]{hyperref}\usepackage{caption}
\usepackage{subcaption}
\usepackage{adjustbox}
\usepackage{enumitem}
\usepackage{longtable}
\usepackage{textcomp}
\usepackage{authblk}

\newcommand{\fbase}{f_{BASE}}

\newcommand*{\revMD}{\textcolor{black}}

\newcommand*{\mac}{\textcolor{black}}

\newcommand*{\macc}{\textcolor{black}}
\newcommand*{\maccc}{\textcolor{black}}

\title{\macc{A first demonstration of active feedback control and multi-frequency imaging techniques for kinetic inductance detectors}}
\author[a]{M. Rouble}
\author[b]{G. Smecher}
\author[a]{M. Adami\v c}
\author[c]{A. Anderson}
\author[d]{P. S. Barry}
\author[e]{K. Dibert}
\author[a,f]{M. Dobbs}
\author[e]{K. Fichman}
\author[a]{J. Montgomery}

\affil[a]{Department of Physics and Trottier Space Institute, McGill University, Montreal, QC, H3A 2T8, Canada}
\affil[b]{t0.technology inc, Montreal, QC, Canada}
\affil[c]{Fermi National Accelerator Laboratory, PO BOX 500, Batavia, IL 60510}
\affil[d]{School of Physics \& Astronomy, Cardiff University, The Parade, Cardiff, CF24 3AA, United Kingdom}
\affil[e]{University of Chicago, 5640 South Ellis Avenue, IL, 60637, USA}
\affil[f]{Canadian Institute for Advanced Research, CIFAR Program in Gravity and the Extreme Universe, Toronto, ON, M5G 1Z8, Canada}

\begin{document}

\maketitle

\begin{abstract}
    \macc{RF-ICE is a signal processing platform for the readout of large arrays of superconducting resonators. 
    Designed for flexibility, the system’s low digital latency and ability to independently and dynamically set the frequency and amplitude of probe tones
    in real time has enabled previously-inaccessible views of resonator behaviour, and opened the door to novel resonator control schemes. We introduce a multi-frequency imaging technique, developed with RF-ICE, which allows simultaneous observation of the entire resonance bandwidth. We demonstrate the use of this technique in the examination of the response of
    superconducting resonators to \maccc{variations in applied readout current and thermal loading}. We observe that, used in conjunction with a conventional frequency sweep at sufficiently large amplitude to induce resonance bifurcation, the multi-frequency imaging technique reveals a resonator response which is not captured by the frequency sweep measurement alone. We demonstrate that equivalent resonant frequency shifts can be achieved using either thermal, optical, or readout loading, and use this equivalence to counteract a change in thermal loading by digitally modulating the readout current through a resonator\maccc{.} We develop and implement a proof-of-concept closed-loop 
    \revMD{negative electro-quasiparticle feedback}
    algorithm which first sets and then maintains the resonant frequency of a lumped element kinetic inductance detector while the loading on it is varied. Although this simple implementation is 
    \revMD{not yet} 
    suitable to deploy at scale, it demonstrates the 
    \revMD{utility  of this feedback technique
    to improve linearity while addressing amplifier distortion, resonator response non-uniformity, and crosstalk. It}
    can be applied to kinetic inductors in non-bolometric operation, and sets the stage for future developments.}
\end{abstract}

\section{Introduction}

As new generations of experiments look to answer new science goals and extend performance, the need for increased sensitivity is important. With established, state-of-the-art detection technologies such as transition-edge sensors (TESs) surpassing the quantum limit of individual sensitivity, further gains are to be made by increasing the number of detectors fielded per instrument. With each detector providing an effectively independent parallel measurement, the observation time required to reach a given sensitivity goal is reduced the more detectors that are operated in parallel. However, achieving these higher detector counts presents certain practical challenges. This is especially true for TESs, whose bolometric nature requires thermal isolation, placing stringent requirements on fabrication capabilities, and whose operation via multiplexing requires additional and often extensive hardware situated at cryogenic stages. \cite{dobbs2012}\cite{irwin2002}\cite{dober2021}

Another superconducting detector technology, the kinetic inductance detector (KID), is gaining popularity in large-scale deployments. While a TES's detection mechanism is an increase in resistance caused by the temperature change associated with photon absorption, a KID is a superconducting resonator whose charge carriers are disrupted by the absorption of energy (either thermal or from incident photons), altering the material's complex impedance and thereby its resonant properties. In essence a resistor, a TES requires additional cryogenic circuitry to enable the multiplexing of the signals from many detectors along one readout line. In contrast, KID resonators can be fabricated in parallel along a feedline on a single wafer, with frequency-domain multiplexing naturally achieved via variation in the resonant frequencies of each device. This dramatic simplification in both fabrication and operation not only allows very high multiplexing factors, but also a high degree of flexibility in their design. KIDs can be made to directly absorb sky photons or to operate as bolometers sensing a temperature change via thermal link with an external absorber, enabling a broad range of detectable wavelengths. The addition of optical elements on the same cryogenic wafer is also possible, such as a broadband optical coupler connected to an array of KIDs in an on-chip spectrometer. \cite{day2003}\cite{wandui2020}\cite{shirokoff2014}

\paragraph{\maccc{Operational challenges}}

These advances in detector design and capability must be accompanied by advances in the technology and techniques used to operate them. In a typical frequency-multiplexing scheme, each detector requires at least one dedicated readout channel, which synthesizes and demodulates a carrier sinusoid at or near its resonant frequency (typically GHz for microwave KIDs). Modern synthesis and signal processing hardware is capable of producing the necessary carrier frequency arrays, but processing constraints still impose limits on the instantaneous bandwidth that a system can achieve.
This in turn drives the design of frequency-multiplexed KID arrays to pack as many resonators as possible into a given readout bandwidth, which can be difficult to reliably achieve. Slight material or fabrication process variations can cause unpredictability in the frequency placement of resonators, which, when many resonators must be densely arranged on a single array, may lead to overlap or collisions between detectors' resonant features. This increases undesirable crosstalk of signals between them and reduces the yield of operable detectors per wafer.

Because their detection mechanism is primarily the change of their resonant frequency in response to changes in absorbed loading, if the responsivities of individual KIDs are not identical, even in the ideal case of identically-varying loading, resonators may drift into each other's influence range, collide, or even cross over one another. This is especially problematic when resonators on an array are deliberately differentially loaded as in a spectrometer, where atmospheric molecular lines may deposit drastically larger loading on some resonators than on others. A system which purely \textit{reads out} resonators, such as by using carrier frequencies which are either static or which vary to track their target resonators, can only mitigate this problem, using thorough characterization of the array responsivities and careful bookkeeping of individual detector positions at all times. \textit{Correcting} this problem requires continuously controlling the frequency placement of all resonators on the array as the loading on them varies, such as by varying the readout power deposited on each resonator in order to keep its total loading effectively constant.

\paragraph{\maccc{Resonator feedback control}}

This control can be achieved passively for thermal kinetic inductance detectors (TKIDs) \cite{wandui2020}, where placing each resonator on a thermally-isolated island allows the dissipation of readout current in the resonator to heat the island. This thermal loading varies as the resonator moves in frequency due to absorbed optical loading, changing the impedance at the frequency of the readout drive tone and modulating the current through it. Similar to the electrothermal feedback which enables most astrophysical TES readout systems \cite{irwin1995}\cite{lee1996}, this technique has been demonstrated in TKIDs to improve the linearity and response time of the detector. \cite{agrawal2021}

Further improvements can be achieved by invoking active, rather than passive, resonator control schemes. Rapidly digitally altering the readout current that is sent to a resonator keeps it locked in place in frequency as the absorbed power on it varies. 
\maccc{When this technique is applied to TKIDs, the gain of the feedback loop can be made much larger than in a passive system, with correspondingly larger improvements in linearity and response time.}
Such improvements would be particularly useful for TKIDs, whose bolometric operation allows them to observe sky frequencies below the superconducting gap energy, but with an inherently long detector time constant. 

However, the fabrication of thermally-isolated islands is complex and reintroduces many of the challenges facing the development of large TES arrays, so we therefore consider whether such a control scheme may be applied to conventional KID arrays.
Non-thermally isolated KIDs are easier to fabricate, and therefore lend themselves more readily to deployment in larger arrays. Due to their integration with the large thermal mass of the silicon wafer, it is likely not practical to control their temperature with the absorption of readout current. However, \mac{we introduce the idea and demonstrate that} it is possible to exert active resonator control on non-thermally-isolated KIDs through the dissipation of readout current 
\revMD{to energize} 
the quasiparticle distribution\macc{.}

We provide a demonstration of this control technique using the RF-ICE readout platform, and describe the hardware, firmware, software, and algorithmic advancements that enabled it.

\section{\macc{RF-ICE}}

RF-ICE is a readout platform designed for high-channel-count, frequency-multiplexed operation of kinetic inductance detectors at GHz frequencies. Developed on the established ICE general-purpose signal processing motherboard, which it pairs with the commercially-available Analog Devices AD9082-FMCA-EBZ mixed-signal front-end daughterboard \cite{ad9082board}, RF-ICE utilizes a modular design to operate up to 1024 superconducting resonators per module, with an arbitrary number of modules operated in parallel. The present generation of this readout framework operates on the ICE-based hardware \cite{Bandura2016} and will deploy on the SPT-SLIM on-chip spectrometer pathfinder experiment. \cite{karkare2021} The readout framework is also compatible with the forthcoming t0 Control and Readout System (CRS) \footnote{\url{www.t0.technology}} RFSoC readout hardware, on which it will deploy as the readout system for the fourth-generation South Pole Telescope experiment, SPT-3G+. \cite{spt2011}\cite{anderson2022} A detailed overview of the ICE-based RF-ICE hardware and firmware can be found in \cite{rouble2022}. This paper focuses on the work using the ICE-based hardware; more detail on its implementation using the CRS RFSoC hardware can be found in \cite{montgomery2024} (these proceedings).

Each RF-ICE motherboard houses two readout modules, each comprising two synthesizers, one demodulator, and an internal pathway for \maccc{low}-latency digital feedback.
Both the synthesizer and demodulator use polyphase filterbanks to generate and digitize carrier sinusoids, with up to 1024 operating per carrier comb. The instantaneous bandwidth spanned by each module is approximately 500 MHz, which can be placed anywhere within a 0 to 3 GHz analog bandwidth using a numerically-controlled oscillator internal to each module.

In operation with superconducting resonators, RF-ICE uses a series of attenuators on the cryogenic stages at the input to the resonator array, and a series of amplifiers (cryogenic and room temperature) on the output line to scale the amplitude of the returning carrier comb to maximize usage of the dynamic range of the ADC. The input-line attenuation reduces the scale of room-temperature Johnson-Nyquist noise to below that of the attenuator immediately preceding the resonators on the coldest stage. The dominant \mac{source of white noise from the readout} is the LNA's input noise. The DAC contributes a 1/f spectrum which is dominant at low frequencies. \mac{The overall system noise is dominated by the detector.}

\subsection{\macc{Resonator characterization}}

Developed alongside prototype detectors for the SPT-3G+ and SPT-SLIM experiments, RF-ICE's control interface provides software routines for many typical aspects of resonator characterization, including resonance finding, parameter fitting, on- and off-resonance noise measurement, and resonator bias point optimization, as well as an extensive bookkeeping framework for mapping physical resonators to digital readout channels for large arrays of resonators.

\begin{figure}[htbp]
    \centering
    \includegraphics[width=0.7\linewidth]{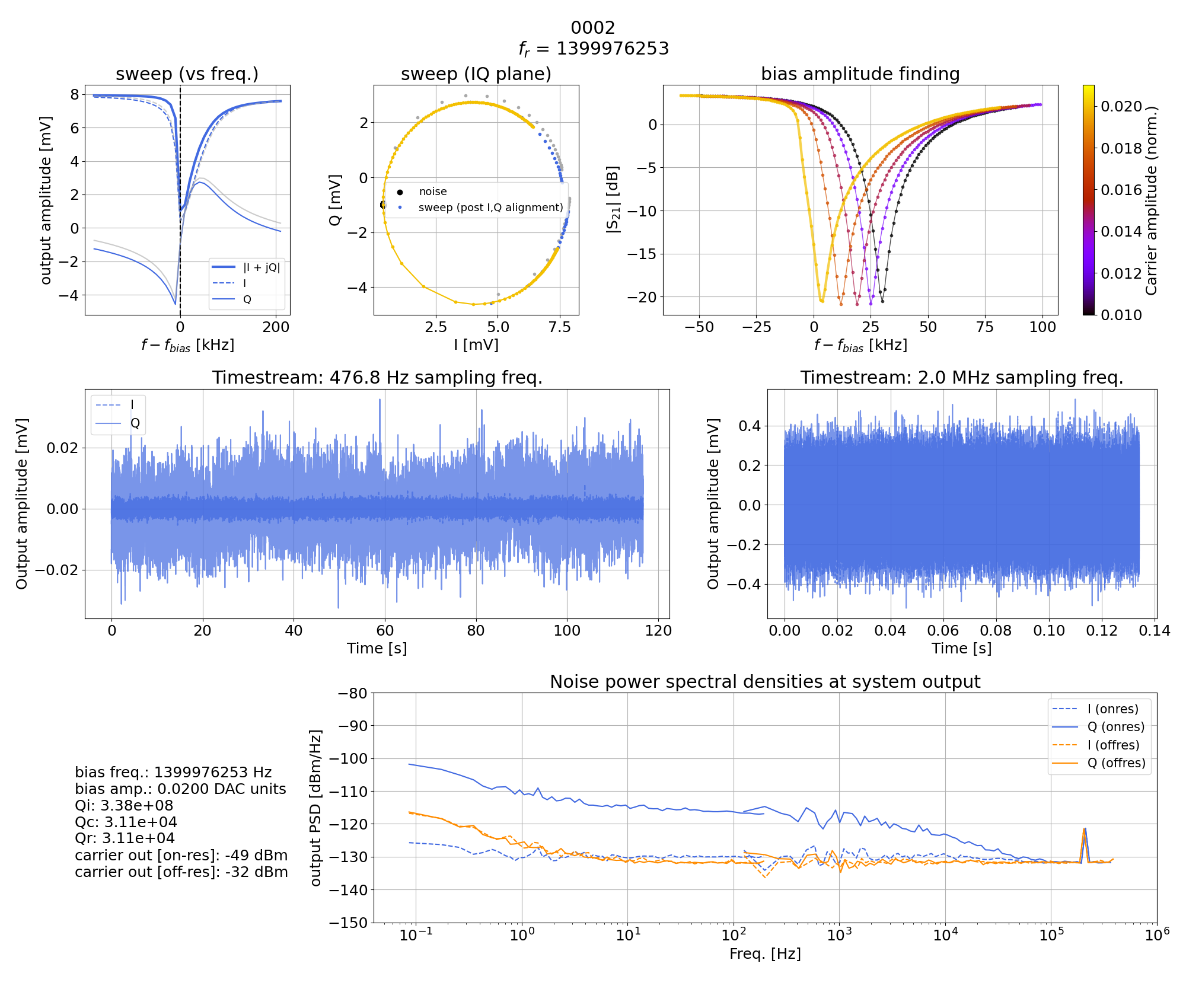}
    \caption{Example output per-detector status digest from the RF-ICE array initialization software algorithm. \maccc{The algorithm performs localized frequency sweep measurements, verifies and updates the bias amplitude and frequency, captures noise timestreams at low and high sampling rates, and activates the carrier comb in preparation for the start of an observation. With the exception of the high sampling rate noise data (which can only be streamed at 2 MSPS from one channel at a time), all operations are performed in simultaneously for all resonators in the array. }
    The gathered data is saved and processed into products useful for both quick-look heads-up displays for individual device debugging (such as above) and for longer-term tracking of array and device performance across multiple observation sessions or time periods.
    }
    \label{fig:resonator_digest_example}
\end{figure}

\macc{RF-ICE uses each module's 1024 independent readout channels to accomplish the detector and array characterization tasks that would traditionally be performed with a network and spectrum analyzer. \maccc{It parallelizes operations across multiple channels, improving efficiency relative to traditional network and spectrum analyzer techniques.} 
This streamlines the characterization of large numbers of resonators in the lab, and allows the rapid initialization of detector arrays at the beginning of an observation. To facilitate commissioning of new arrays and instruments, the software routines produce a wide range of data products and heads-up displays, organized at the level of the individual resonator and of the array as a whole. An example resonator-level display is shown in Figure \ref{fig:resonator_digest_example}, featuring a collection of data products captured by the array initialization algorithm. The algorithm captures this data for all resonators in the array simultaneously (excepting the high sampling rate noise timestream which is taken one channel at a time), then collates them into various representations and formats for preliminary inspection and further analysis. Shown here for an example detector, these displays provide readily-accessible summaries of system status for the user and enable rapid debugging.}

\subsection{Multi-probe: RF-ICE multi-frequency imaging mode}

To implement the measurement routines described in the previous section, \mac{RF-ICE uses an operational mode which is akin to a system of network and spectrum analyzers, operated in parallel.}
Each measurement in the routines is performed by synthesizing and recording \mac{carriers, each at a given frequency. In this `traditional' mode, m}easurements spanning a frequency range are made sequentially in time as well as in frequency, as in a network analyzer. \maccc{Measurement sets are performed in parallel across multiple channels.}

A multi-channel readout platform such as RF-ICE can \mac{re-imagine} this paradigm and measure multiple frequency points across a range simultaneously, rather than sequentially. This allows the user to capture a snapshot of the system's behaviour at every frequency of interest at a given moment in time, rather than seeing only the frequency of the probe as it sweeps across. We refer to this technique as a \textit{multi-probe} measurement\mac{. This simple change in operational mode can reveal a surprising wealth of previously-inaccessible information about a given system.}

\begin{figure}[htbp]
    \centering
    \includegraphics[width=0.5\linewidth]{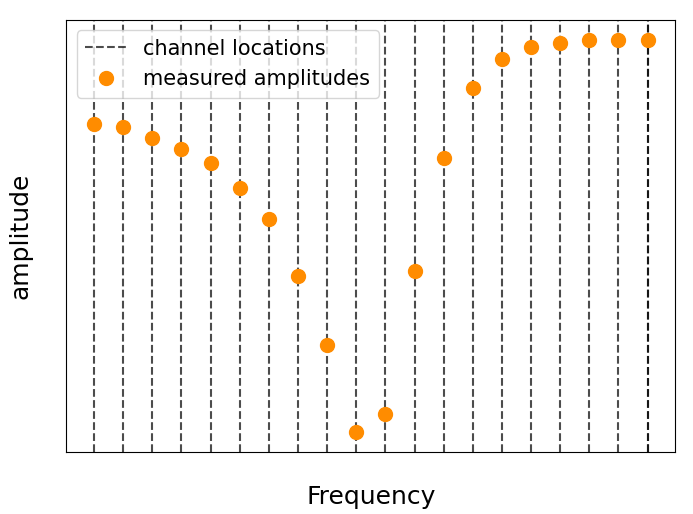}
    \caption{Illustration of the multi-probe measurement technique. `Watcher' channels at very low amplitude are spaced across some bandwidth of interest (here, across a resonator). A single data capture is taken from all channels simultaneously, and the measured voltages (orange points) are displayed versus the watcher channel frequencies (indicated by dashed lines). This technique allows the imaging of the entire resonator's response at a given moment in time, as opposed to the conventional frequency sweep technique, in which the displayed data is sequential in both frequency and time.}
    \label{fig:multiprobe}
\end{figure}

Figure \ref{fig:multiprobe} shows \mac{an example} of a resonator imaged using a multi-probe measurement, as a demonstration of the technique. A number of very low amplitude `watcher' channels are placed within a bandwidth of interest. The positioning of the channels in frequency space is arbitrary, but they are typically placed approximately evenly across the bandwidth. In order to not perturb the system under test, the amplitudes of the watcher channels must be kept very small, generally at least a factor of 100 smaller than a typical readout probe tone. On all devices studied for this work, it was experimentally verified that at the default small amplitude, the presence of the watcher tones had no measurable effect on the resonance.
Once the watcher channel frequencies are set, a batch of samples is captured from all channels in the module simultaneously through the standard RF-ICE data acquisition pathway. Each channel's data are averaged into a single complex voltage value, which is saved to disk along with the channel frequency. Fig. \ref{fig:multiprobe} shows an array of such values, plotted as the absolute value of the complex I,Q voltage pairs versus the channel frequencies at which they were acquired. In this way, the multi-probe measurement captures an arbitrary bandwidth at a single moment in time.

This is of particular use for observing superconducting resonators at high readout probe amplitudes, as it is well-documented that the absorbed power of the probe affects the resonator's behaviour. \cite{swenson2013}\cite{deVisser2010}\cite{goldie2013} \maccc{I}n traditional network analyzer mode, the user sees the system at the probe frequency only, and thus the final measurement product convolves the response of the resonator to the probe current with the motion of the probe itself across the resonance. This leads to \mac{an incomplete picture of the resonator's dynamics, with the resonator exhibiting} complicated nonlinear behaviour which is often challenging to model analytically \mac{from within this framework}.

\begin{figure}[!htbp]
    \centering
    \begin{subfigure}[c]{\textwidth}
        \centering
        \includegraphics[width=0.4\textwidth]{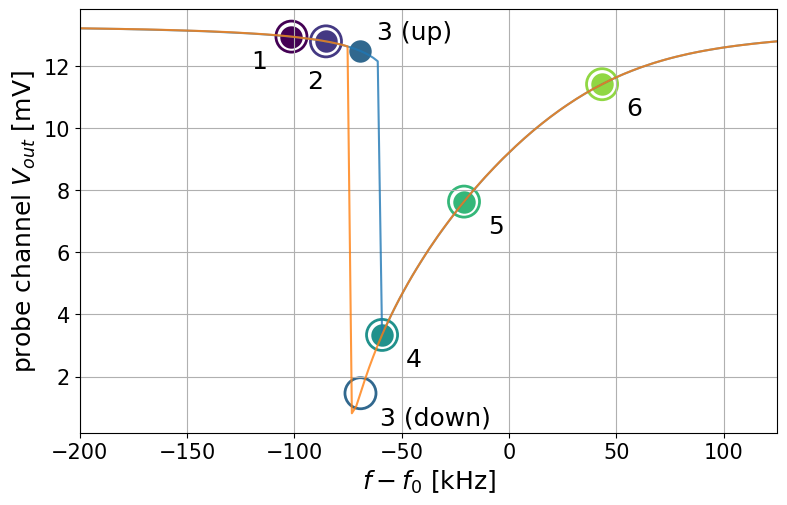}
        \caption{Frequency sweep measurements of a resonator, in ascending (blue) and descending (orange) frequency order\mac{, made using a traditional network analyzer-like single-tone probe}. The amplitude of the probe channel, whose measured output voltage amplitude is plotted here versus its frequency in each step of the measurement, is sufficiently large to trigger characteristic bifurcation behaviour in the resonator. The frequency sweep was performed in discrete steps, with the probe frequency held stationary at each frequency step for 1-2 seconds before jumping to the next frequency. At each frequency step, with the probe frequency constant, a multi-probe measurement (Figure \ref{fig:sweep_multiprobe_view}) of the same resonance was captured. A subset of these multi-probe measurements is plotted in Fig. \ref{fig:sweep_multiprobe_view}, as individual frames corresponding to the indicated probe frequency.}
    \label{fig:sweep_probe_view}
    \end{subfigure}

    \begin{subfigure}[c]{\textwidth}
    \centering
        \includegraphics[width=0.7\textwidth]{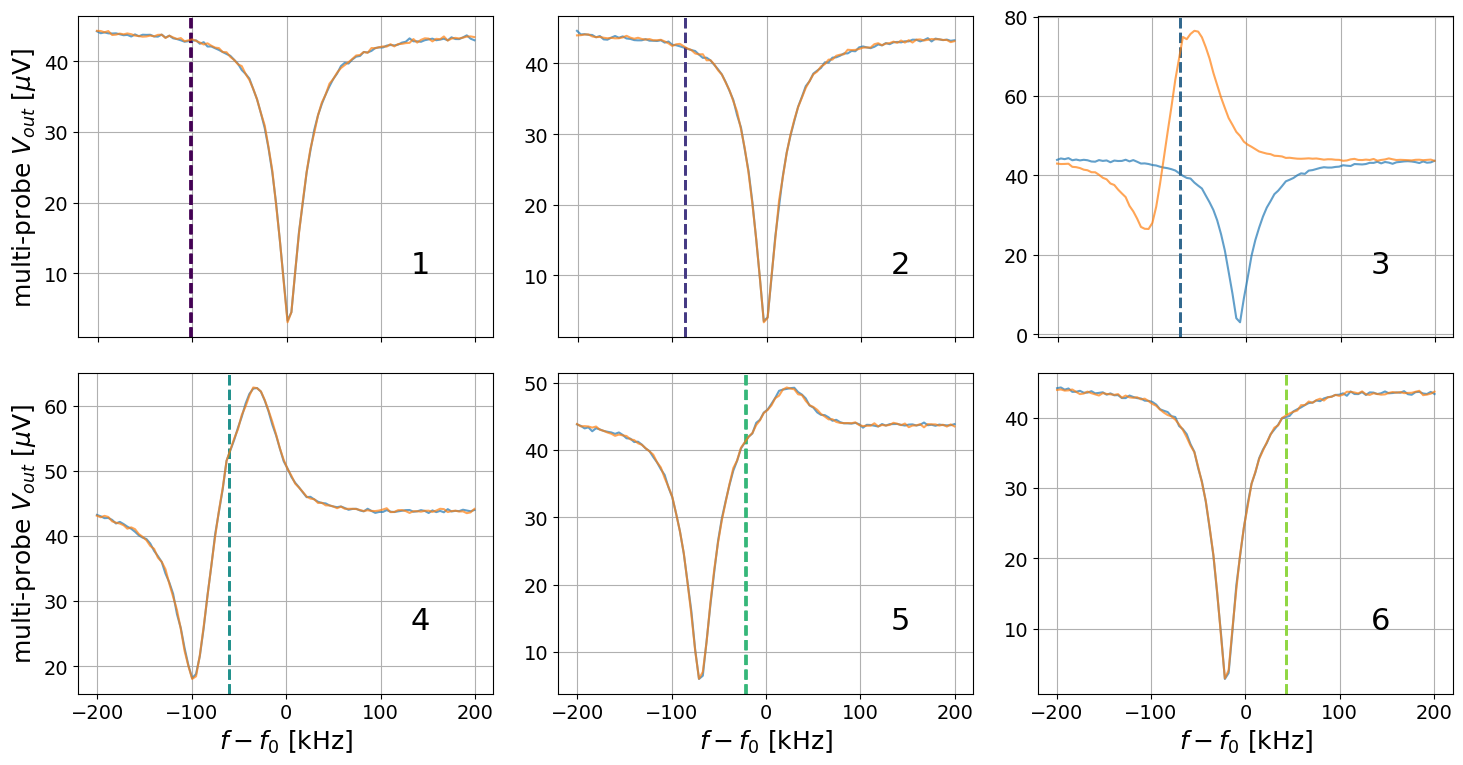}
        \caption{Multi-probe measurements of the same resonance as in Figure \ref{fig:sweep_probe_view} made during the same frequency sweep measurement. The panels are ordered by time, with each corresponding to one of the probe frequency steps plotted in Fig \ref{fig:sweep_probe_view}. Upwards and downwards frequency sweep directions are indicated by the same colours as in \ref{fig:sweep_probe_view}.  The location of the probe channel is indicated in each panel by a vertical dashed line. The probe itself \mac{
        \revMD{is not shown}
        in this view}, as it does not enter within the bandwidth of any of the watcher channels. Measured voltages from the watcher channels trace the shape of the resonance and reveal its response to the stimulus of the probe tone as it is swept across the bandwidth. The discontinuity which is characteristic of bifurcation in a traditional sequential frequency sweep measurement is here seen as the sudden shape change in the upward sweep between panels 3 and 4, as the resonance snaps from its typical inverted Lorentzian shape into a distorted state with some watcher channels experiencing positive gain. On the downward sweep (reading the panels in descending order), the onset of the distortion is gradual\maccc{: in panel 6, the resonance is displays a frequency shift due to probe current flowing through it. As the probe moves further into the resonance,} the magnitude of the effect grows until panel 3 when the resonance snaps back into its regular state in panel 2 (the discontinuity seen in the downward sweep). As expected from the traditional frequency sweep measurements, the jump between the two states occurs at a lower frequency on the downward sweep than the upward, pushing the resonance further into the distorted state than is accessible on the upward sweep. Note that the voltage values reported by the probe channel \textit{do not} trace the same shape reported by the watcher channels -- the probe channel does not observe the positive gain, and instead measures a further decreased output signal amplitude where the largest positive gain occurs in the watchers. In this state, the driven resonator is parametrically amplifying the watcher channels. The exploitation of parametric amplifiers is presently an active field of research, and \macc{so the appearance of this behaviour in resonators intended for use as kinetic inductance detectors may be of interest as a medium for further study.}}
    \label{fig:sweep_multiprobe_view}
    \end{subfigure}
\end{figure}

\maccc{This distinction is well-illustrated by performing a frequency sweep measurement, while using additional channels to monitor the entire resonance bandwidth.}
Figure \ref{fig:sweep_probe_view} shows the traditional network analyzer view of a probe sweeping across a resonator at sufficient amplitude to trigger typical bifurcation behaviour, such as is characterized in detail in \cite{swenson2013}. The resonance exhibits a hysteric dependence on the direction of the probe sweep. The multi-probe view (Figure \ref{fig:sweep_multiprobe_view}), obtained by placing a number of small-amplitude watcher channels at frequencies spanning the resonance while the single large-amplitude probe channel is swept across, provides a much greater amount of information about the processes underway.

Each panel of \ref{fig:sweep_multiprobe_view} corresponds to one measurement position of the probe tone (indicated by a vertical dashed line and corresponding to one of the numbered points in \ref{fig:sweep_probe_view}). On the upward sweep, as the probe moves toward the resonance, we see the resonance shift to slightly lower frequency as the readout current through it increases. The eventual abrupt transition (occurring between panels 3 and 4 in the upward sweep and 3 and 2 in the downward sweep) corresponds to the bifurcation discontinuity seen in the probe view, although unlike in the conventional frequency sweep view, here the resonance is seen to snap into a distorted or excited state, with the watcher channels receiving positive gain. As the probe moves out of the resonance on the other side, the excited state relaxes back to the usual resonance curve.

\mac{Notably}, the shape traced by the probe channel does not match the shape reported by the watcher channels. Indeed, the lowest amplitude sections of the probe's data correspond to the frequency steps at which the watcher channels receive the highest gain. At these probe frequencies, the amplitude of the probe current flowing through the resonator is sufficient for parametric amplification of the watcher channels to become dominant, in a manner which is qualitatively reminiscent of the 4-wave mixing described in \cite{chien2023}. \mac{The multi-probe view has allowed us to directly observe and report on this remarkable behaviour in these devices.}

\maccc{This multi-frequency view of resonator hysteretic switching behaviour suggests that} the probe channel's amplitude is distributed amongst the nearby watcher channels, increasing their amplitudes while decreasing that of the probe itself. As the probe moves toward the resonance, the amount of current flowing through it increases. This increase is compounded by the motion of the resonance toward the probe frequency due to the nonlinear increase in kinetic inductance. This positive feedback eventually causes the resonance to jump into the excited state, which is seen as \mac{an apparent} bifurcation in the sequential frequency sweep measurement. \cite{swenson2013} On the downward sweep, the probe not only pushes the resonance away as it moves toward it, but also loses amplitude to the distortion of the resonance and to the watcher channels. This causes it to appear to trace out a deeper part of the resonance.

A deeper treatment of this phenomenology will be presented in a later work. 
\mac{For the purposes of this work, we use this new multi-probe capability to measure the detectors' response to light, temperature, and readout power (Section \ref{sec:resonator_control_theory}), and, in Section \ref{sec:active_resonator_control}, develop a new control scheme via \macc{negative electro-quasiparticle feedback} for kinetic inductance detectors.}

\section{Resonator control via readout-quasiparticle excitation}\label{sec:resonator_control}
\subsection{Resonator circuit model under optical, thermal, and readout current loading}\label{sec:resonator_control_theory}

The absorption of heat, light, and readout power impacts the number and distribution of quasiparticles in a superconducting resonator. In steady state, the density of quasiparticles is constant, so the rates of generation from all sources are balanced by the rate of recombination:

\begin{equation}
    \frac{dn_{qp}}{dt} = 0 = \Gamma_{th} + \Gamma_{opt} + \Gamma_{read} - \Gamma_r
\end{equation}

Solving this for $n_{qp}$ then allows the estimation of the complex conductivity $\sigma = \sigma_1 - j\sigma_2$ of the superconductor \cite{gao2008}:

\begin{equation}
    \sigma_1 = \sigma_N \frac{2 \Delta_0}{h f} \frac{n_{qp}}{N_0 \sqrt{2 \pi k_B T \Delta_0}} K_0(\xi)
\end{equation}

\noindent and
\begin{equation}
\sigma_2 = \sigma_N \frac{\pi \Delta_0}{hf} [1 - \frac{n_{qp}}{2N_0\Delta_0} (1 + \sqrt{\frac{2\Delta_0}{\pi k_B T}}e^{-\xi} I_0(\xi))]
\end{equation}

\noindent where $\sigma_N$ is the normal-state conductivity for the sample, $I_0$ and $K_0$ are modified zeroeth-order Bessel functions of the first and second kind respectively, $\xi = \frac{hf}{2k_B T}$, $\Delta_0$ is the zero-temperature gap energy, $h$ is the Planck constant, $N_0$ is the single spin density of states for aluminum, and $k_B$ is the Boltzmann constant.

From this with the geometry of the superconductor, we can compute the kinetic inductance and dissipative impedance via the surface impedance \cite{henkels1977}\cite{deVisser2014}:

\begin{equation}
    Z_s = \sqrt{\frac{j 2 \pi f \mu_0}{\sigma}} (\mathrm{tanh}(t \sqrt{j 2 \pi f \mu_0 \sigma}))^{-1}
\end{equation}

\noindent where $t$ is the thickness of the superconductor, $\mu_0$ is the permittivity of the material, $f$ is the frequency at which the impedance is being evaluated, and $\sigma$ is the complex conductivity.

Putting all this together, we obtain values for the resistance and kinetic indutance as

\begin{equation}
    R = \mathrm{Re}(Z_s) \frac{l}{w} \quad \text{and} \quad L_k = \frac{\mathrm{Im}(Z_s)}{2 \pi f} \frac{l}{w}
\end{equation}

\noindent (where $l$ and $w$ are the length and width of the inductor) based on the expected density of quasiparticles in a resonator under a given set of loading conditions.

\begin{figure}
    \centering
    \includegraphics[width=0.7\linewidth]{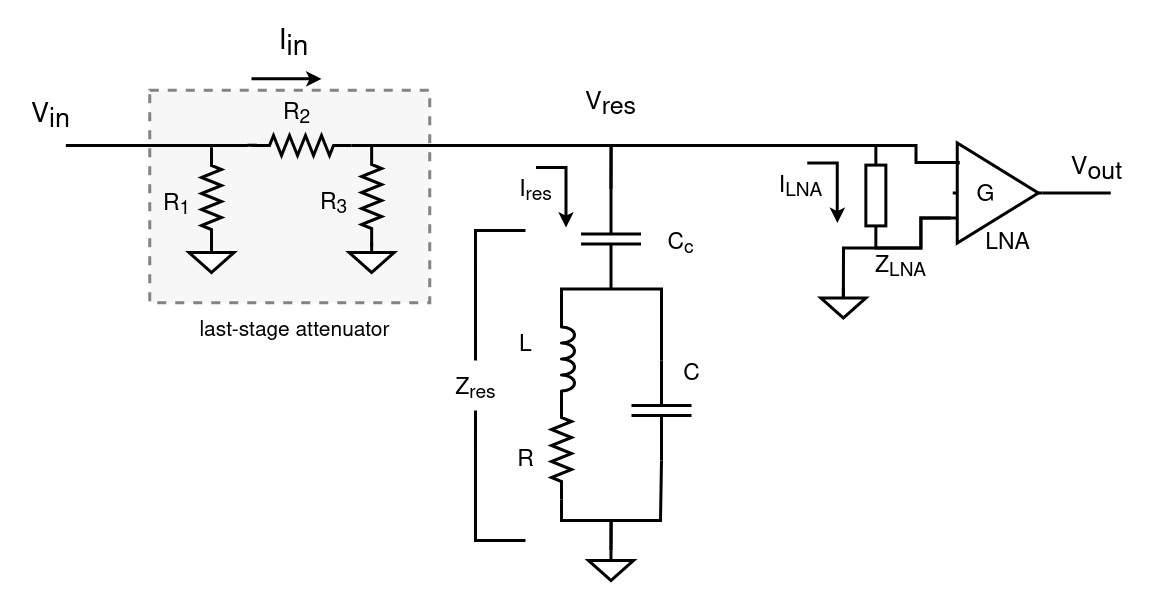}
    \caption{Circuit model of a single lumped element resonator and surrounding readout electronics. On resonance, the impedance of the resonator is small but non-zero (typically $\lesssim 10 \Omega$), and purely real. Readout drive current amplitude varies widely between resonator designs, and is difficult to calculate precisely due to the many frequency-, geometry-, and temperature-dependent impedances in the system.
    For the resonators studied in this work, estimates of current amplitude through the resonator are on the order of 0.1-1 $\mu$A, giving an approximate dissipated power on-resonance of $\lesssim$ pW. This is comparable to the expected sky loading, and suggests that modulating the readout drive current can effectively compensate for a modulation in sky loading.}
    \label{fig:circuit_model}
\end{figure}

The devices studied in this work are lumped-element KIDs, with an aluminum inductor whose superconducting impedance we derive above, and a physical planar capacitor (either interdigitated or parallel-plate) which is typically niobium and which is assumed to not contribute to loading-dependent effects. \cite{dibert2022}\cite{dibert2023}\cite{robson2024}\cite{barry2022} The resonator is coupled to the feedline with a small coupling capacitor. A circuit model for this type of resonator in its environs within the readout system is shown in Figure \ref{fig:circuit_model}. The resonator (or array of resonators) is in parallel with the input impedance of the cryogenic LNA, and forms the load on the last-stage cryogenic attenuator.

A complete solution of the impedance of the system involves a frequency- and geometry-dependent simulation for the specifics of each detector on the wafer and the nearby electrical connections, in order to account for signal reflections and propagation effects. Nonetheless, considering the lumped elements in a standard circuit analysis provides useful intuition and a reasonable degree of accuracy. In this framework, we model the resonator as a inductance (comprised of the kinetic component computed as above and a geometric component which does not change with loading) with small series resistance (arising from the real component of the complex conductance), in parallel with the lumped element capacitor:

\begin{equation}
    Z_{RLC} = \big[ (j2\pi f L + R)^{-1} + (\frac{1}{j2\pi f C})^{-1}\big]^{-1}
\end{equation}

\noindent all of which is in series with the coupling capacitance $C_c$ such that the total resonant branch has impedance

\begin{equation}
    Z_{res} = \frac{1}{j 2\pi f C_c} + Z_{RLC}
\end{equation}

\noindent The transfer function for this system from the point labeled $V_{in}$ to the output of the LNA is:

\begin{equation}\label{eq:transfer_function}
    V_{out} = V_{res} G_{LNA} = V_{in} \frac{R_3 || (Z_{res} || Z_{LNA})}{(R_3 || (Z_{res} || Z_{LNA})) + R_2} G_{LNA}
\end{equation}

To understand the impact of a change in loading (whether from optical, thermal, or readout dissipation), we evaluate the change in quasiparticle density, and propagate this through to a measurable change in the transfer function. The resonant frequency is a function of all four lumped elements in the resonator, but can be approximated as $2 \pi f_r = (L C)^{-1/2}$, making it easy to see that an increase in the kinetic inductance will result in a decrease in resonant frequency.

The response of superconducting resonators to optical and thermal loading is well-studied (see, for example, \cite{gao2008equivalence}, \cite{gao2008}, \cite{zmuidzinas2012}) and will not be discussed here in detail.
The impact of readout power absorption on the quasiparticle population has been studied in several works (\cite{deVisser2010}\cite{goldie2013}\cite{deVisser2014}, among others). Readout photons generally have energies much lower than the superconducting gap energy of aluminum, and therefore cannot individually directly break Cooper pairs into quasiparticles. There is nonetheless substantive readout current dissipation in a resonator, which affects the charge carrier distribution.

\maccc{The impact of readout power absorption on a resonator may be described using a model of a driven, non-equilibrium distribution which may be approximated to considerable accuracy as an equilibrium thermal distribution at elevated effective temperature, $T_{eff}$. \cite{goldie2013}\cite{thomas2015}}
In this framework, we may expect that loading a resonator via readout power absorption will result in a frequency shift analogous to that incurred by increased thermal loading.

\begin{figure}
    \centering
    \includegraphics[width=0.4\textwidth]{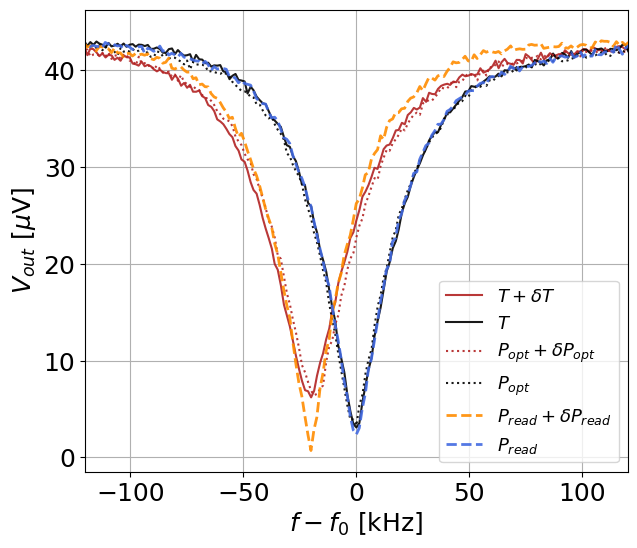}
    \caption{A resonator's response, measured using the multi-probe technique, showing equivalent frequency shifts when loading on it is varied by either a temperature change (solid lines), an optical loading change (dotted lines), or by varying the amplitude of a readout tone at fixed frequency (dashed lines).  The relative shift of the device's resonator frequency is well matched between the different loading sources. Unlike increased loading from thermal or optical sources, an increased drive current generally results in an enhanced resonance Q factor\maccc{, possibly} due to saturation of TLS \cite{pappas2011}, an effect which is particularly prominent on this device.}
    \label{fig:dT_dP_dLED}
\end{figure}

This equivalence is demonstrated in Figure \ref{fig:dT_dP_dLED}, which compares the response of a resonator under a change in thermal and optical loading to its response after altering the amplitude of a drive tone (at fixed frequency) to achieve the same frequency shift. In the device shown, the increased readout current has the additional effect of increasing the resonator's quality factor, possibly due saturation of two-level systems (TLS), as described in \cite{pappas2011}.
\mac{The ability of a modulation in the readout current to counteract a frequency shift due to loading change shows that readout probe power is viable for negative 
\revMD{electro-quasiparticle}
feedback, to lock a detector's resonant frequency in place as sky signal changes.}

In thermally-isolated KIDs operating as bolometers, the dissipation of readout current can be used as \mac{passive electrothermal feedback (ETF) to linearize the detector response.}\cite{agrawal2021} In this architecture, \mac{which resembles ETF in transition-edge sensor bolometers \cite{lee1996},} the dissipation contributes to the total thermal load on the KID, made possible by its isolation from the bulk of the substrate. As the resonance moves in response to thermal load from absorbed optical load, the changing impedance at the drive tone frequency (located \textit{above} the resonant frequency) modulates the amount of current flowing through and readout power dissipated in the resonance. This elegant negative feedback stabilizes the resonant frequency of the device, improving its linearity and response time.

The devices studied in this work are direct-absorption KIDs and are not thermally isolated from the \maccc{wafer substrate}. We therefore \mac{expect that} the temperature of the device does not meaningfully differ from its surroundings, \mac{so we cannot employ the same ETF scheme described above. Instead,} control of the resonant frequency occurs through the excitation of the quasiparticle distribution, rather than true thermal loading.

To illustrate this, consider a slightly modified rate equation for the total population of quasiparticles in the system: 
\begin{equation}
    \frac{dn_{qp}}{dt} = \Gamma_{T_{eff}} + \Gamma_{opt}  - \Gamma_r
\end{equation}

where $T_{eff}$ is the effective temperature of the driven quasiparticle distribution, as described in \cite{goldie2013}. $T_{eff}$ is a function of the physical temperature of the superconductor (the bath temperature $T_b$) and the absorbed power per volume in the resonator, with higher powers leading to higher effective temperatures, and $T_{eff} > T_b$. A change in the rate of quasiparticle production from optical sources, $\Gamma_{opt}$, may therefore be countered by the opposite change in $\Gamma_{T_{eff}}$, which may be achieved via modulation of the readout current through the resonator.

\begin{figure}
    \centering
    \includegraphics[width=\linewidth]{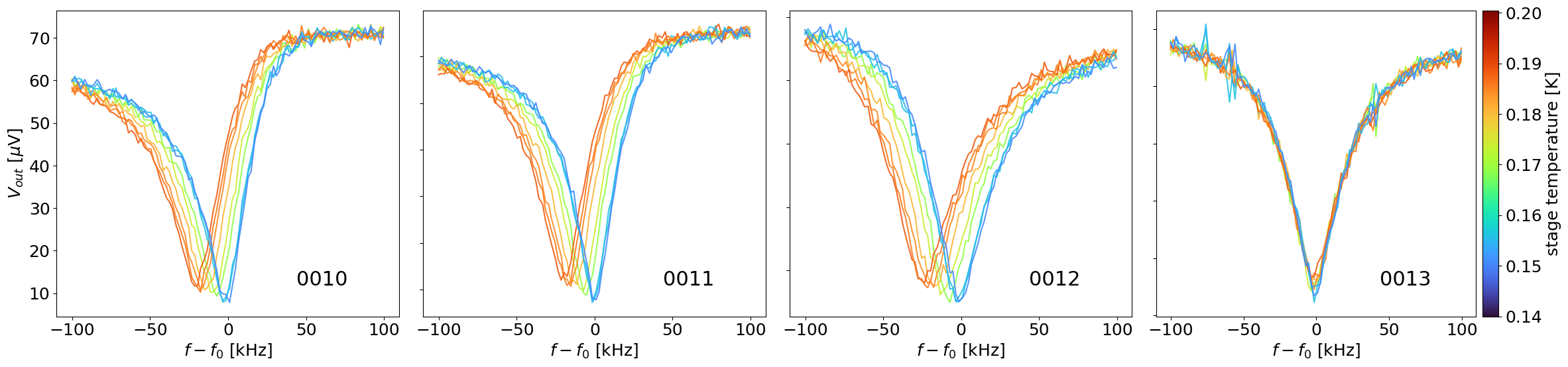}
    \caption{Proof-of-concept demonstration of stabilizing the resonant frequency of a single resonator via readout-quasiparticle excitation during a cryogenic stage temperature change\mac{, which we use as an easy-to-control proxy for incident optical power variation}. A fixed-frequency drive tone is placed 200 kHz above the resonant frequency of resonator 0013. The stage temperature is increased, which causes a negative frequency shift in the other, undriven resonances, but which is counteracted by decreasing the amplitude of the drive tone on 0013. As a result, 0013 remains at the same resonant frequency, while the other resonances drift to lower frequencies. The applied drive tone amplitudes are recorded, and can be used to reconstruct the applied temperature change. All four resonances are imaged here using simultaneous multi-probe captures taken every few seconds throughout the temperature sweep. Although under realistic operational conditions this control \maccc{would} be achieved using a single channel per resonance to both monitor position and apply drive control, for this simple demonstration, these multi-channel measurements of 0013 were used to digitally adjust the drive tone amplitude \mac{in feedback} to keep its resonant frequency static.}
    \label{fig:slow_temperature_sweep}
\end{figure}

In Figure \ref{fig:slow_temperature_sweep}, we demonstrate that this technique \mac{works} by modulating the amplitude of a readout drive tone at static frequency\mac{, using a software algorithm that implements feedback,} in order to counteract the effect of a temperature change on a resonator. Other resonators in the array which are not controlled in this way exhibit a shift to lower frequency as the temperature of the wafer is increased, but the controlled resonator remains fixed in place. 

This demonstration serves as a proof-of-concept for \mac{an active} quasiparticle excitation control technique \mac{that can be used for KIDs, whether or not they are thermally-isolated}. However, the utility of the open-loop control algorithm implemented in the demonstration is limited, \maccc{as it is dominated by networking delays and} makes use of multiple readout channels per resonance in order to monitor the resonance's position and \maccc{calculate} the drive tone amplitude. In a deployment context, this must be achieved \maccc{at low latency and} using a single readout channel to both monitor and control the behaviour of the target resonator. An RF-ICE-based early-stage implementation of such a closed-loop active control system using quasiparticle excitation is described in Section \ref{sec:active_resonator_control}.

\subsection{Active Resonator Control}\label{sec:active_resonator_control}

\mac{The new} active resonator control (ARC) system \mac{presented here} uses a digital feedback mechanism to modulate the applied readout drive current in response to a changing signal on a detector. This fixes the device's resonant frequency in place under conditions where it would otherwise drift to a lower or higher value as the loading on it \mac{is} changed. It can be applied to TKIDs in bolometric operation (true heating), or non-thermally-isolated resonators (quasiparticle excitation).

\mac{A detailed} discussion of feedback theory will be covered in a separate work. Here we present a phenomenological overview of an early-stage implementation of a closed-loop ARC system using quasiparticle excitation, its use to stabilize resonators under temperature \mac{or incident optical loading} changes, and discuss motivating factors for the development of such a system.

\subsubsection{\macc{Implementation}}

The control system will strive to keep the total \mac{density of quasiparticles} in the resonator roughly constant, and so the resonator must first be initialized to an elevated density before beginning the feedback control loop. This initialization involves the placement of a drive tone at a frequency at a chosen offset below the un-biased resonant frequency of the target device. The amplitude of the tone is increased, increasing the current through the resonator and driving up its effective quasiparticle distribution temperature until the resonant frequency \mac{is pulled in to reach} the drive tone. The optimization of the initial placement of the drive tone with respect to the resonance, and its starting amplitude before ramping up, \mac{depend on the properties of the resonator and the expected signal power changes.} 
The initialized amplitude sets the dynamic range of the feedback, and so must be sufficient to compensate for the anticipated range of variations in signal loading.

Once the resonator is initialized, the control loop begins. The final complex output voltage from the initialization routine is saved and kept as the target value, to which the feedback will lock the system. At each iteration of the loop, the drive channel measures a voltage and computes an error term with respect to the target voltage. This error term is passed into a proportional controller, which determines the new drive amplitude. Because the measured voltage is directly proportional to the drive amplitude (as can be seen in to the transfer function for the circuit given in Section \ref{sec:resonator_control_theory}), either the target value or the measured value must be scaled by the drive amplitude at each iteration.

\begin{figure}[!htbp]
    \centering
    \includegraphics[width=0.8\textwidth]{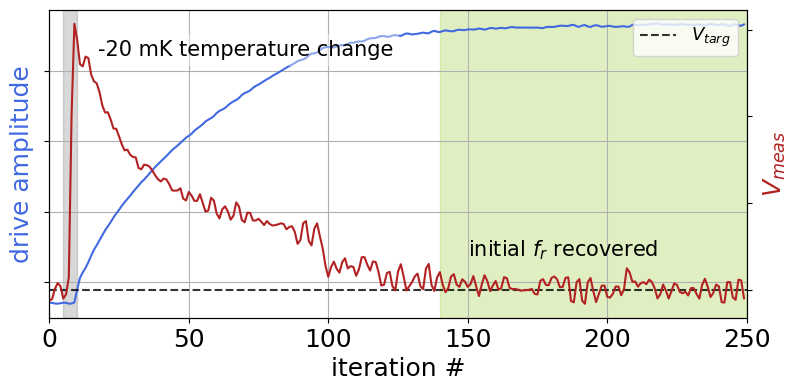}
    \caption{Proof-of-concept demonstration of closed-loop active resonator control, using a single readout channel to stabilize the resonant frequency of KID when the cryogenic stage temperature (again used as a simple proxy for optical loading variation) is abruptly lowered 20 mK from its starting point.
    In this implementation, the feedback controller resides in software on a control PC, and so the control loop is severely limited by variable delays in network communication to and from the RF-ICE readout hardware. The system is nonetheless able to recover the starting resonant frequency of the target resonator by modulating the drive amplitude sent into the system, and feeding back on measurements of the output carrier voltage, $V_{meas}$, in comparison with the target voltage, $V_{targ}$, established by the initialization routine.
    Because in this implementation the control lags the response of the resonator, this system is analogous to slow or intermittent tone-tracking, albeit with the advantage that the carrier frequencies can be fixed to avoid in-band IMD
    \revMD{and maintain uniform spacing to avoid collisions and variations in crosstalk}. 
    Future implementations will use a \maccc{low}-latency digital feedback pathway onboard the readout hardware itself, reducing the latency of the controller below the response time of the detector and allowing it to exert much greater control over the resonance. In this case, the resonant frequency no longer varies under changing loading, and the primary science signal is instead in the feedback drive current.}
    \label{fig:slow_feedback_demo}
\end{figure}

A proof of concept implementation of this technique is shown in Figure \ref{fig:slow_feedback_demo}. In this example, the controller is implemented \mac{in software} on the readout control computer, and so the latency and instability of the loop are large and strongly dominated by communication over the network to the RF-ICE board. Nonetheless, in the example shown, the controller is able to recover the starting resonant frequency of the target resonance after a 20 mK stage temperature change, while uncontrolled resonances in the array experienced an average frequency shift of 4 kHz. Future implementations\mac{, under development now,} will move the control loop into 
\revMD{low-latency FPGA firmware, so that it can operate across the full
detector bandwidth.}

\subsubsection{Discussion}

\macc{Generally speaking, the designers of a readout system operating KIDs must choose where to 
place carrier tones with respect to the resonant frequencies of the detectors and have a strategy for keeping them at the chosen point, as deviations from the ideal bias frequency will result in reduced sensitivity of the output voltage to changes in signal on the detector. They must also choose the amplitude with which to bias the resonators, which scales the signal amplitude relative to system noise sources, and devise a way to mitigate dynamical effects within the array, such as increased overlap or collisions between neighbouring resonances in response to differential loading.}

\paragraph{\macc{Intermodulation distortion in a passive readout system}}\label{sec:carrier_frequencies}

\revMD{Every electronic system includes elements with nonlinearities. This leads to intermodulation distortion (IMD) products
that increase with the number and amplitude of carriers that are multiplexed together
in a detector module.}
IMD occurs when signals at 
\revMD{two or more}
frequencies pass through a nonlinear element, such as an amplifier, and results in the re-distribution of signal power to non-signal frequencies. Of particular concern are the odd-order IMD products, which tend to fall near the signal frequencies.
\macc{To maximize sensitivity over system noise sources, kinetic inductance detectors typically use large carrier amplitudes to bias the detectors. As the impact of IMD scales with both the number of carriers in the system and with the amplitude of these carriers, care must be taken when operating high-multiplexing-factor KID arrays to avoid spectral contamination and/or reduced sensitivity.}

In a frequency-multiplexed readout system, which necessarily involves \mac{the summing} of many frequencies along a single conductive path, care must be taken to mitigate the impacts of IMD. IMD arises from the \revMD{nonlinear mixing} of frequencies produced by a wide variety of sources in the system, \revMD{such as} the assigned frequency of each carrier in the comb, the sample rate of the ADC, the IceBoard's upconverted clock frequency, the switching frequency of buck converters on the IceBoard and the AD9082 mezzanine, and the frequency precision of the direct digital synthesizer (DDS).

\macc{When operating as a passive readout system,} RF-ICE adopts a mitigation strategy for IMD products which places all third-order IMD products outside the 
\revMD{several-hundred}-\mac{hertz-wide} bandwidth \revMD{occupied by the sky signal in} every readout channel 
during operation. This is accomplished by ensuring that all frequencies in the system are integer multiples of a base frequency greater than the \revMD{sky signal} 
bandwidth. The chosen base frequency for RF-ICE is $\fbase \simeq 476.8$ Hz. All other key frequencies listed above are integer-divisible by this number, which is itself integer-divisible by the DDS frequency precision, which sets the smallest achievable difference between two frequencies for the system.

\revMD{With this strategy, } IMD products occur at integer multiples of $\fbase$ and land either on top of the carriers themselves or outside the channel bandwidth. Carrier frequencies may be regularly updated (in integer steps of $\fbase$) to track the position of a target resonance.
Without this frequency scheduling, the positions of the carriers are arbitrary in relation to one another. When few carriers are active, the arbitary placement and base frequency aligned cases are equivalent, as there are so few IMD spurs produced that it is unlikely that one will fall within a given carrier bandwidth. At higher active channel counts however, the arbitrary placement case introduces a very large number of IMD products at small individual amplitudes, which contribute an excess noise to the system. \maccc{This excess can then be deliberately whitened with an additional modulation step, trading narrowband spectral contamination for a general increase in white noise level}. \cite{yu2023}

\macc{In operation, loading changes result in shifts in the resonant frequency of the detectors. The \maccc{user} has the choice of continuously following the resonance by updating the carrier frequency as needed to maintain a constant bias point with respect to the resonance (tone tracking), intermittently updating the carrier locations to maintain a proximity to the resonant frequency (quasi-static), or leaving the carriers in place (true static). In the static and quasi-static cases, carrier frequencies can maintain base-frequency alignment by quantizing any frequency adjustments made to integer multiples of $\fbase$. This alleviates the risk of in-band IMD and so is not negatively impacted by the use of large carrier amplitudes, but is at the expense of being continually slightly detuned from the resonance. When tone tracking, the carriers may be maintained at the ideal tuning point, but IMD products may appear at any frequency and are made larger by the use of larger carrier amplitudes.} 

\macc{None of these strategies can prevent collisions (two resonances overlapping) in the presence of differential loading changes. In a tone-tracking system, differential loading is also accompanied by changes in carrier spacing, resulting in dynamically changing crosstalk and IMD spurs which may be difficult to account for in analysis.}
\macc{Choosing between these strategies requires a detailed understanding of the science targets of an experiment and the characteristics of the observations needed to achieve them. The motivation is also clear for the exploration of new strategies which might allow both continuous optimal tuning and large carrier amplitudes, without spectral contamination from IMD.}

\paragraph{\revMD{Benefits of ARC}}

The ability to control the resonant frequencies of the devices in an array offers advantages which are apparent even in this early stage of development. In a densely-packed array, it can prevent collisions between nearby resonators which either have different responsivities or which are receiving different amounts of loading, by locking them to their assigned frequencies. 
It also allows the user some freedom of choice in the selection of the precise resonant frequencies for all the detectors in the array, as this is determined by the application of the drive tones over a range of 10s of kHz (as we have demonstrated here), and perhaps further. While it is not likely that this technique can be used to disentangle resonances which are significantly overlapping in their undriven state, this may still allow some relaxation of fabrication constraints when designing dense arrays. Further, this allows the readout system to maintain a base-frequency carrier schedule with the resonances `pulled' by the readout current to perfectly match the carrier placements, as described above.
\revMD{For thermal KIDs,}
passive electrothermal feedback can stabilize the location of the resonance in frequency space. \revMD{However,} the technique requires that the drive tone be located above the resonant frequency in order to achieve negative feedback, and so by definition cannot maintain the ideal bias frequency.

In a system where \mac{noise from a source such as the cryogenic LNA is not negligible on-resonance,} increasing the amplitude of the carrier tones increases the system's sensitivity. The amplitude of carrier tones in an active control system are typically larger than those achievable in a passive system, as the resonators are being biased at or beyond the onset of nonlinearity seen in traditional sequential frequency sweep measurements. Given the additional headroom above such a noise source, a system employing active control might choose to relax its requirements on LNA noise performance, and/or choose a unit with increased dynamic range, which would in turn reduce concerns about IMD.

Although the simple, software-based ARC demonstrated here is too slow to be used in a practical observational context, it demonstrates the possibility of using digitally-modulated readout current to control and improve the operation of kinetic inductance detectors. In a fast, strong feedback regime such as is widely employed on TESs in astronomical observations, the response time of the detector is reduced as a function of the feedback loop gain, widening its useful science bandwidth. \cite{lee1996} Further, the feedback linearizes the device response, and extends its dynamic range. The development of such a fast active control system, building on the principles demonstrated here
\revMD{and implemented using the existing firmware modules employed for digital feedback in transition edge sensors} \cite{montgomery2021}\cite{deHaan2012} is currently in progress, and will be reported on in a later work.

\section{Conclusion}

The ambitions of future experiments to field ever-higher numbers of detectors and in increasingly creative designs are motivating \maccc{the growing adoption} of kinetic inductance detectors in astrophysics and cosmology. A promising but still comparatively new detector technology to be used at scale, considerable work is currently underway to address the challenges these KID deployments face and improve their performance on the sky.

The development of new tools and techniques is an important part of this process. With the introduction of the multi-probe imaging technique, we have reported on the remarkable behaviour of these superconducting resonators in their response to absorbed readout power.
The picture this measurement technique provides of the interaction of these resonators with the dissipation of readout current is much more complete than what can be obtained using conventional sequential frequency sweep measurements. It remains in good agreement with existing models of inductive nonlinearity as a non-equilibrium excitation of the quasiparticle distribution, and suggests parametric conversion of the large probe tone into the amplitudes of secondary observer tones as a novel explanation for the hysteretic switching behaviour which is well-documented in these devices.

With this new technique and the insights it has provided us, we have demonstrated the ability to exercise resonant-frequency control over kinetic inductance detectors without the need for thermal isolation. Although still in an early phase of development for these devices, this type of feedback control was a key development in the large-scale use of transition-edge sensors for millimeter-wavelength cosmology.
Future implementations of the basic feedback control algorithm demonstrated here will move the controller on-board the readout hardware, where the ICE/CRS low-latency digital feedback pathway will greatly expand its gain and bandwidth. This holds the promise of improved response time, dynamic range, system linearity, and sensitivity for future systems fielding this type of control architecture on large arrays of kinetic inductance detectors.

\section{Acknowledgements}

The McGill authors acknowledge funding from the Natural Sciences and Engineering Research Council of Canada, \revMD{the Canadian Foundation for Innovation,} and Canadian Institute for Advanced Research.
\revMD{The detectors used in this work} made use of the Pritzker Nanofabrication Facility of the Institute for Molecular Engineering at the University of Chicago, which receives support from Soft and Hybrid Nanotechnology Experimental (SHyNE) Resource (NSF ECCS-2025633), a node of the National Science Foundation’s National Nanotechnology Coordinated Infrastructure.

\bibliography{bibliography}
\bibliographystyle{spiebib}

\end{document}